\documentclass[11pt,a4paper]{article}
\usepackage{jcappub}
\usepackage{comment}

\newcommand{\cg}{{\cal G}}
\newcommand{\cK}{{\cal K}}

\begin{document}

{\baselineskip0pt
\rightline{\baselineskip16pt\rm\vbox to-20pt{
           \hbox{YITP-16-59, KUNS-2582}
\vss}}%
}

\title{
Bigravity from gradient expansion
}

\author[a]{Yasuho Yamashita}

\author[a,b]{Takahiro Tanaka}

\affiliation[a]{Yukawa Institute for Theoretical Physics,
Kyoto University, 606-8502, Kyoto, Japan}
\affiliation[b]{Department of Physics, Kyoto University, 606-8502, Kyoto, Japan}

\emailAdd{yasuho@yukawa.kyoto-u.ac.jp}
\emailAdd{t.tanaka@tap.scphys.kyoto-u.ac.jp}


\arxivnumber{1510.07551}

\abstract{
We discuss how the ghost-free bigravity coupled with a single 
scalar field can be derived from a braneworld setup. 
We consider DGP two-brane model without radion stabilization. 
The bulk configuration is solved for given boundary metrics, and 
it is substituted back into the action to obtain the effective 
four-dimensional action.
In order to obtain the ghost-free bigravity, we consider the gradient expansion 
in which the brane separation is supposed to be sufficiently small 
so that two boundary metrics are almost identical. 
The obtained effective theory is 
shown to be ghost free as expected, 
however, the interaction between two gravitons 
takes the Fierz-Pauli form at the leading order of the gradient expansion, 
even though we do not use the approximation of linear perturbation. 
We also find that the radion remains as a scalar field in the 
four-dimensional effective theory, but its coupling to the 
metrics is non-trivial.  
}

\maketitle
\flushbottom

\section{Introduction}
In general, bigravity, {\em i.e.}, the gravitational model that contains two gravitons
interacting with each other, 
was known to be suffered from an unavoidable ghost mode, which 
is called Boulware-Deser (BD) ghost~\cite{BD}.
Recently, however, the restriction of the interaction to the specific form 
of non-derivative coupling proposed by de Rham, Gabadadze and Tolley
(dRGT) in massive gravity~\cite{dRGT1, dRGT2}
is found to evade the BD ghost problem also in the ghost-free bigravity \cite{HR1, HR2, Hassan:2011ea}. 
This ghost-free bigravity allows us to investigate many applications 
of bigravity to building a cosmological model, 
which revealed that 
the ghost-free bigravity can be consistent with the observation 
of the current universe~\cite{Comelli:2011zm, Comelli:2012db, DeFelice:2013nba, Fasiello:2013woa, DeFelice:2014nja, Konnig, otherBG}, 
although there exists some constraints on the model construction 
to avoid the gradient instability in cosmological perturbation 
at the early epoch~\cite{Comelli:2012db, DeFelice:2014nja,
Konnig, Aoki}. 
As a challenge to the viability of the model 
there are also discussions about the 
superluminality and acausality pointed out in the context of massive gravity \cite{Deser:2012qx, Deser:2013eua,
Deser:2013qza, Deser:2014fta},
or the stability of the tuning of the ghost-free interaction 
against radiative corrections~\cite{deRham:2013qqa, Heisenberg:2014rka}.

The interaction between two metrics 
different from the dRGT form must be suppressed to realize a 
viable model, 
but we do not know the mechanism that realizes 
such fine tuning. 
In order to find a clue to this fine-tuning problem and 
also to give a UV completion of the ghost-free bigravity model, 
we attempted to derive the ghost-free bigravity as a low-energy effective 
theory starting with a healthy braneworld setup, which is naturally 
expected to be free from BD ghost~\cite{Yamashita:2014cra}.
In order to derive bigravity from the braneworld setup, 
we have to realize the mass hierarchy 
between the lowest massive Kaluza-Klein (KK) 
mode and the other KK modes.
We considered Dvali-Gabadadze-Porrati (DGP) 2-brane model \cite{DGP}, 
because the Einstein-Hilbert terms localized on the branes effectively 
trap graviton modes on the respective branes, 
and as a consequence only the lowest and the second lowest masses 
of KK towers of gravitons, which correspond to these two localized
modes, can be made to be light by tuning the contributions of the 
brane localized Einstein-Hilbert terms to be 
large compared with the bulk one~\cite{Padilla, Yamashita:2014cra}. 
This tuning is achieved by setting $\Delta y/ r_c^{(\pm)}\ll1$, 
where $\Delta y$ is the brane separation and 
$r_c^{(\pm)}:=M_{pl}^{(\pm)2} / 2M_5^3$ 
with $M_{pl}^{(\pm)}$ and $M_5$ being the 4-dimensional 
Planck masses for the brane localized Einstein-Hilbert terms
and 5-dimensional Planck mass, respectively. 
An alternative way to lead to bigravity by discretizing the extra-dimension
has been investigated in Ref.~\cite{deRham:2013awa}. 

In this paper, we do not introduce any radion stabilization
mechanism which fixes the brane separation, 
and hence there exists a scalar radion corresponding to the degree of
freedom of the brane separation. 
Namely, the low-energy effective action should consist of 
two gravitons and one scalar radion. 
This model is even simpler to analyze because we do not have to employ
the stabilization mechanism. 
Nevertheless, since the radion couples to both induced metrics on branes, 
there is a possibility that the low-energy effective 
theory derived from DGP 2-brane model may naturally give 
a non-trivial ghost-free doubly-coupled matter model 
different from the previously proposed ones~\cite{Yamashita:2014fga, deRham:2014naa, Noller:2014sta, deRham:2014fha,Huang:2015yga, Heisenberg:2015iqa, Heisenberg:2015wja}.
Here, we should recall that DGP 2-brane model has two branches: 
the self-accelerating branch inevitably has 
either the Higuchi ghost or the radion ghost, 
while the normal branch is free from ghost 
when de Sitter universe is investigated~\cite{Charmousis, Izumi}. 
In order to get the ghost-free branch, 
we need to set the model parameters to satisfy 
$1-2r_c^{(\pm)}(n_{(\pm)}^\mu \partial_\mu \log a)> 0$ 
on the respective branes~\cite{Yamashita:2014cra}, 
where $a$ is the background bulk warp factor and $n_{(\pm)}^\mu$ 
is the unit normal vector pointing toward the bulk on each brane. 
If we assume that the values of the extrinsic curvature 
on both branes are approximately the same, {\em i.e.}, 
$K\approx K_+\approx K_- \approx \mp 4(n_{(\pm)}^\mu \partial_\mu \log a)$,  
the above condition requires $|K|\lesssim 1/ r_c^{(\pm)}$. 
Combining the condition $\Delta y/ r_c^{(\pm)}\ll1$, 
the region of the model parameters of our interest
is restricted to 
\begin{equation}
\left| K\Delta y \right| \ll 1\,.
\end{equation}
Since our interest is in the regime with small $K \Delta y$, 
we use the gradient expansion~\cite{Kanno:2002iaa} to construct the 
bulk solution. 
We can systematically expand the action with respect to 
the small quantity, $K\Delta y$.  
Since the higher-order terms in the gradient expansion cannot cancel the
lower-order ones, we naively expect that 
the action at each order is kept free from ghost. 
In this paper we consider the lowest order 
truncation of the gradient expansion. 
In order to obtain the action of the low-energy effective theory 
written in terms of the metrics on the $(\pm)$-branes, $g_{\mu\nu}^{(\pm)}$, 
we solve the bulk equations for given boundary metrics $g_{\mu\nu}^{(\pm)}$
at the lowest order of the gradient expansion, and then 
we integrate out the bulk degrees of freedom 
by substituting back the obtained bulk solution into the action 
and performing the integration along the extra dimension.  
Our method using the gradient expansion will not always give a
ghost-free effective action even if we start with a healthy 
higher dimensional model 
because the higher order corrections in the gradient expansion 
would naturally derive higher-derivative terms in the effective action 
which seem to introduce extra degrees of freedom including ghost modes 
in addition to the expected ones, which are two graviton modes 
and one scalar radion in the present case. 
We should interpret that the appearance of such extra degrees of freedom  
is responsible for the other bulk degrees of freedom with higher masses, 
which are in the present setup 
the massive KK gravitons with the mass squared of $O(\Delta y^{-1})$ 
or higher. Therefore, the appearance of extra degrees of 
freedom itself will not directly indicate the breakdown of the gradient 
expansion. Although it would be unsatisfactory if such higher 
derivative terms remain even in the final result of the low-energy 
effective action, they should be understood to be treated by means of 
the method of order reduction, then. 

Besides the fine-tuning problem of the ghost-free interaction form, 
the extension of the ghost-free bigravity to braneworld setup may solve 
the issue of gradient instability in cosmology and acausality, 
since both of them seem to be absent in the braneworld setup. 
The main difference between bigravity and the brane model 
will originate from the presence of the extra 
massive graviton modes in the braneworld setup, 
which are inactive at low energies but will 
play a crucial role at high energies.
Also, we expect that the stability against radiative corrections
might be understood well from the viewpoint of the braneworld
setup. However, our study presented in this paper is not matured yet to 
address these issues. 

This paper is organized as follows: In Sec.~II, we introduce the braneworld
model that we consider in this paper and explain 
the strategy to obtain the effective action. 
In Sec.~III, we start with the derivation of the effective action 
to the quadratic order in perturbation around a de Sitter brane
background solution, and show that the obtained effective action 
contains the expected two gravitons and one scalar radion. 
In Sec.~IV, we analyze the equation of motion  
derived from the effective action obtained in Sec.~III with source
energy momentum tensor
and investigate the positivity of the action when we integrate out 
all the gravitational field degrees of freedom. 
In Sec.~V we extend the method to obtain the effective action to the nonlinear level. 
In Sec.~VI, we summarize the result and discuss possible future extension of
the current work.

\section{Model}
We consider DGP two-brane model without the radion stabilization, whose action is written as
\begin{eqnarray}\label{DGP2brane}
 S = S_b + \sum_{\sigma=\pm} S_\sigma \,,
\end{eqnarray}
where 
\begin{eqnarray}
 S_b&:=& \frac{1
}{2} \frac{M_{pl}^2}{2r_c^{(+)}} \int d^5x \sqrt{-{}^5\!g} 
 \left({}^5\!R - {12\over \ell^2_\Lambda}\right)\,, \\
 S_+&:=&  \frac{M_{pl}^2}{2} \int d^4x \sqrt{ -g_{(+)}} \left( R_{(+)} - 2\sigma_+ \right) \,, \\
 S_- &:=& \frac{\chi M_{pl}^2}{2} 
\int d^4x \sqrt{ -g_{(-)}} \left( R_{(-)} - 2\sigma_-  \right)\,,
\end{eqnarray}
where $\!^{5}\!g_{\mu\nu}$, $\!^{5}\!R$, $g_{\mu\nu}^{(\pm)}$, $R_{(\pm)}$ 
are the 5-dimensional metric, 
5-dimensional Ricci scalar, 4-dimensional metrics and Ricci scalars 
induced on $(\pm)$-branes, respectively.
$M_{pl}$ is the 4-dimensional Planck mass on $(+)$-brane, 
while $M_{pl}^2/2r_c^{(+)}$ and $\chi M_{pl}^2$ are  
the 5-dimensional bulk Planck mass cubed and 4-dimensional Planck mass
squared on the $(-)$-brane, respectively.
$\sigma_\pm$ are the 4-dimensional cosmological constants 
on the respective branes, 
while $6/\ell^2_\Lambda$ is the 5-dimensional cosmological constant. 
The equation of motion in the bulk spacetime is given as 
\begin{eqnarray}\label{bulkeom}
 \!^{5}\!R_{\mu\nu} -{1\over2}\, \!^{5}\!R \,^{5}\!g_{\mu\nu} + {6\over \ell_{\Lambda}^2}=0 \,.
\end{eqnarray}
Here, we assume the $Z_2$ symmetry across each brane. 
We adopt $S_1/Z_2$ orbihold identification where the left hand 
side and right hand side with respect to each brane are identical. 
Then, the extrinsic curvatures on the $(\pm)$-branes, $K_{\mu\nu}^{(\pm)}$, are 
determined by the Isra\"el's junction conditions:
\begin{eqnarray}
 \label{junction}
K^{(\pm)}_{\mu\nu} - K^{(\pm)} g^{(\pm)}_{\mu\nu} 
= \pm r_c^{(\pm)} \left(G_{\mu\nu}^{(\pm)}+\sigma_\pm g_{\mu\nu}^{(\pm)} \right)\,,
\end{eqnarray}
where $r_c^{(-)}:=r_c^{(+)}\chi$. 

We will derive the low-energy effective action from \eqref{DGP2brane},
by solving the bulk metric for given boundary metrics on the branes 
and integrating out the bulk degrees of freedom. 
In this paper, we employ the gradient expansion to find the bulk metric, 
assuming that the brane separation 
is sufficiently small compared with the 5-dimensional curvature scale, 
and hence the metrics on the both branes are identical 
at the leading order of the small separation expansion.
Here, we describe the strategy to obtain the effective action in more detail.
At the lowest order of the expansion, we can set the 
following ansatz:
\begin{equation}
\label{spacetime}
ds^2=N^2(x) dy^2+
g_{\mu\nu} dx^\mu dx^\nu\,,
\end{equation}
while the $(\pm)$-branes locate at $y=y^{\pm}$, respectively. 
We use the gauge degrees of freedom to choose 
the lapse $N$ and the shift $N^\mu$ in the above form. 
Namely, we imposed $\partial_y N=0$ and $N^{\mu}=0$.
For given boundary geometries specified by 
$g^{(+)}_{\mu\nu}(x) dx^\mu dx^\nu$ and 
$\tilde g_{\mu\nu}(\tilde x) d \tilde{x}^\mu d \tilde{x}^\nu$ 
($g^{(-)}_{\mu\nu}(x)$ is reserved for later use) 
on the respective branes, 
we should be able to find an appropriately interpolating bulk solution 
by solving the bulk equation of motion. 
Decomposing the bulk equations \eqref{bulkeom} 
with respect to the $y={\rm constant}$ hypersurfaces
under the above gauge fixing condition \eqref{spacetime}, 
we obtain 
\begin{eqnarray}
  \partial_{\tilde{y}} K_{\mu\nu} & = & -2 K^\rho_\mu K_{\rho\nu} + K K_{\mu\nu}+{4\over
 \ell_{\Lambda}^2}g_{\mu\nu} - R_{\mu\nu} + {N_{;\mu \nu} \over N} \,,
\label{KdotEq}
\\ 
 K_{\mu\nu} &= &-{1\over 2}\partial_{\tilde{y}} g_{\mu\nu}\,, 
 \label{gdotEq}
 \\
 K^2-K^\mu_\nu K^\nu_\mu & = & -{12 \over \ell_{\Lambda}^2} + R\,,
\label{HconstEq}
\\
 K_{;\mu}- K^\nu_{\mu ;\nu} & = & 0\,,
 \label{mconstEq}
\end{eqnarray}
where we define $\partial_{\tilde{y}} \equiv {1\over N} \partial_y$.
$K_{\mu\nu}$ and $R_{\mu\nu}$ are the extrinsic curvature 
and the 4-dimensional Ricci tensor evaluated 
on each $y$-constant surface, respectively. 
Here, we stress that $g_{\mu\nu}$ and $K_{\mu\nu}$ are 
defined not on the $\tilde{y} \equiv Ny={\rm constant}$ hypersurface, 
but on the $y={\rm constant}$ one. 
$f_{;\mu}$ is the covariant derivative of the field $f$ 
associated with the 4-dimensional metric $g_{\mu\nu}$ 
(We reserve $\nabla_\mu$ for later use). 
The indices are raised and lowered also using $g_{\mu\nu}$. 

Once we fix the way how 
to relate the 4-dimensional coordinates on one brane with those on the other 
by the gauge choice of the bulk metric, 
the coordinates on the other brane is specified. 
If we adopt the above bulk metric ansatz \eqref{spacetime}, 
the way of connecting between two branes is completely specified. 
In these specific coordinates, 
the metric is given by 
\begin{equation}
  g^{(-)}_{\mu\nu}(x)= 
  {\partial \tilde x^\alpha\over \partial x^\mu}
  {\partial \tilde x^\beta\over \partial x^\nu}
   \tilde g_{\alpha\beta}(\tilde x)\,, 
\end{equation}
and $g^{(-)}_{\mu\nu}(x)$ cannot be chosen arbitrarily. 
If we solve Eqs.~\eqref{KdotEq}-\eqref{HconstEq} (excluding 
\eqref{mconstEq}) with the boundary metrics 
$g_{\mu\nu}^{(+)}$ and $g_{\mu\nu}^{(-)}$, the coordinate transformation $\tilde x^\alpha(x)$ is 
constrained by the momentum constraint~\eqref{mconstEq}.
Using the gradient expansion,  
we can expand $g_{\mu\nu}$ around the middle point $\bar{y}=0$ in the bulk as
\begin{eqnarray} \label{expand}
 g_{\mu\nu}^{(\pm)} =  \bar{g}_{\mu\nu} + \overline{\partial_{\tilde{y}} g_{\mu\nu}} \tilde{y}^{\pm} 
 + \frac{1}{2} \overline{ {\partial_{\tilde{y}}}\,^2 g_{\mu\nu}} (\tilde{y}^{\pm})^2 + \cdots\,,
\end{eqnarray}
where $\bar{g}_{\mu\nu} := g_{\mu\nu}(0)$, and $\tilde{y}^\pm \equiv N(x) y^\pm$. 
As we set $\bar{y}=0$, the values of $\tilde y$ on the respective 
branes are related with each other as $\tilde{y}^{-}=-\tilde{y}^+$.
Since the positions of the $(\pm)$-branes in $y$ are fixed at $\mp y_0$, 
respectively, we have 
\begin{equation}
  \tilde y^+=  - N y_0<0\,,\qquad 
  \tilde y^-= N y_0>0\,.
\end{equation} 
Using the expression \eqref{expand} 
and the evolution equations \eqref{KdotEq} and \eqref{gdotEq}, 
we can obtain $\bar{g}_{\mu\nu}$ and $\bar{K}_{\mu\nu}$ in terms of  
$g_{\mu\nu}^{(\pm)}$ and $\tilde{y}^+$, 
and then the Hamiltonian constraint \eqref{HconstEq} determines 
$\tilde{y}^+$ and hence $N$ in terms of $g_{\mu\nu}^{(\pm)}$.
The effective action written in terms of $g_{\mu\nu}^{(\pm)}$ should be given 
by substituting the bulk solution into the action 
\begin{eqnarray}
\label{effac}
 S &=& {M_{pl}^2\over2r_c^{(+)}} \oint d^5x \sqrt{-g} 
 \left(R + K^2-K^\mu_\nu K^\nu_\mu - {12 \over \ell_{\Lambda}^2} \right) \nonumber \\
 &&+ {M_{pl}^2 \over 2}\int d^4x \sqrt{-g_{(+)}} \left( R_{(+)} - 2\sigma_+ \right) 
 + {\chi M_{pl}^2 \over 2}\int d^4x \sqrt{-g_{(-)}} \left( R_{(-)} -
						     2\sigma_- \right)
 \, ,
\end{eqnarray}
where the 5-dimensional Einstein-Hilbert action is also decomposed into the 4+1 form 
with respect to the $y={\rm constant}$ hypersurfaces. 
Since the integrand of Eq.~\eqref{effac} does not 
include any second derivative of the metric in the $y$-direction, 
the boundary Gibbons-Hawking terms on the branes are unnecessary in this expression.
The integral along $y$-direction reduces to integrals of 
simple powers of $y$ by expanding the integrand at $y=0$
with the aid of the gradient expansion.
We do not manifestly impose the momentum constraints, 
because they are automatically imposed later. 
As long as the extrinsic curvatures evaluated on the branes, 
$K_{\mu\nu}^{(\pm)}$, for the obtained bulk solution, 
agree with the ones that are derived from 
the variation of the effective action with respect to
$g_{\mu\nu}^{(\pm)}$, as is expected, 
we obtain the junction conditions \eqref{junction}
from the variation of the effective action. 
By taking the divergence of the junction conditions \eqref{junction}, the right hand side 
identically vanishes, and hence we find that the momentum constraints \eqref{mconstEq}
are imposed. 

\section{Perturbation around de Sitter spacetime}
First, we consider the perturbation around a de Sitter brane solution with the comoving curvature $H$ 
and calculate the effective action quadratic in the perturbation,
as a warm-up.
The background spacetime is given by
\begin{eqnarray}
 ds^2 &=& dy^2 + a^2(y) \gamma_{\mu\nu}dx^\mu dx^\nu\, , 
\end{eqnarray}
where we set $N=1$, $a(y=0)=1$ and $\gamma_{\mu\nu}$ 
is the 4-dimensional de Sitter metric with the expansion rate $H$. 
Then the background solution is determined by the Hamiltonian constraint \eqref{HconstEq} 
and the junction condition \eqref{junction}:
\begin{eqnarray} 
 \label{Hge1}
\mathcal{H} &:=& {\partial_y a \over a} = 
\pm\sqrt{-{1 \over \ell_{\Lambda}^2} + {H^2\over a^2}} \, , \\
\label{jcge1}
\mp \mathcal{H}_\pm&=&  r_c^{(\pm)} \left( -{\sigma_{\pm} \over 3} + {H^2\over a_\pm^2} \right) \, ,
\end{eqnarray}
where $\mathcal{H}_\pm:=\mathcal{H}\left( y=\mp y_0 \right)$ 
and $a_\pm:=a\left( y=\mp y_0 \right)$. 
As we choose the convention that ${\cal H}(y=0)>0$, 
the $(+)$-brane should have a positive tension ($\sigma_+>0$). 
Then, Eq.~\eqref{Hge1} is solved as
\begin{eqnarray}\label{mathcalH}
 \mathcal{H} = -{1\over \ell_{\Lambda}} \tan{\left( {y\over
					       \ell_{\Lambda}}+A\right)}\,, 
\end{eqnarray}
which is the background value of $-K/4$, 
with the integration constant 
\begin{eqnarray}
 A&=&-\arccos\left({1\over  \ell_{\Lambda}H}\right)  \left( <0 \right) \,.
\end{eqnarray}
Then we obtain 
\begin{eqnarray}\label{a}
 a = B \cos{\left( {y\over \ell_{\Lambda}}+A\right)}\,, 
\end{eqnarray}
with the integration constant 
\begin{eqnarray}
 B&=& \ell_{\Lambda}H \,.
\end{eqnarray}
Using the above expression, we find 
\begin{equation}
 \mathcal{H}_\pm:=
 \mathcal{H}\left(\mp y_0 \right)=
 {1\over \tilde{\ell}_\Lambda} 
 \pm H^2 y_0 +{1\over \tilde{\ell}_\Lambda} \mathcal{O}(H^2 y_0^2).
\label{Hpmexpand}
\end{equation} 
where we define $\tilde{\ell}_\Lambda$ by
\begin{equation}
 \tilde{\ell}_\Lambda^{-1}:= \mathcal{H}(y=0)
 =\sqrt{-\ell_{\Lambda}^{-2} + H^2} \,.
\end{equation}
We represent the perturbation around this background as 
\begin{eqnarray}
 g_{\mu\nu}^{(\pm)} &=& a^2(\mp y_0) \left( \gamma_{\mu\nu} + h_{\mu\nu}^{(\pm)} \right) \, , \\
 N&=&1+x\,, \quad \mbox{\em i.e.}\,, \quad \tilde{y}^\pm = \mp y_0 (1+x)\, .
\end{eqnarray}
Here we note that the value of $y$-coordinate at the location of 
the perturbed $(+)$-brane remains to be $y=-y_0$ in our coordinate system. 
To make the calculation easier, we define new variables as
\begin{eqnarray}
\label{cgdef}
 {\cg}_{\mu\nu}\left( y \right) &:=& a^{-2}(y) g_{\mu\nu}( y)\,, \\
 {\cK}_{\mu\nu}\left( y \right) &:=& a^{-2}(y)\left(K_{\mu\nu} (y)
 + {\mathcal{H}(y)\over N} g_{\mu\nu}(y) \right)\,,
\end{eqnarray} 
so as to isolate the $y$-dependence related to the 
background warp factor $a(y)$. 
Then, with these new variables, Eqs.~\eqref{KdotEq}-\eqref{mconstEq} become
\begin{eqnarray}
  \partial_{\tilde{y}} \cK_{\mu\nu} & = & -2 \cK^\rho_\mu \cK_{\rho\nu} 
  + \cK \cK_{\mu\nu} - {\mathcal{H}\over N} 
\cK \cg_{\mu\nu} 
  - 4 {\mathcal{H}\over N}  \cK_{\mu\nu}  \nonumber \\
  &&
   + \left( {4 \over \ell^2_\Lambda} + 4{\mathcal{H}^2\over N^2}  
  +    {\partial_{\tilde{y}} \mathcal{H}\over N} \right) \cg_{\mu\nu}
  - a^{-2}R_{\mu\nu} + a^{-2}{ N_{;\mu\nu} \over N} \,,
\label{tKevolEq} 
\\
\cK_{\mu\nu} &= &-{1\over 2}\partial_{\tilde{y}} \cg_{\mu\nu}\,, \\
\label{tHconstEq}
 \cK^2 - 6 {\mathcal{H}\over N}  \cK - \cK^\mu_\nu \cK^\nu_\mu 
 & = & R - 12\left( {\mathcal{H}^2\over N^2} 
+ {1\over \ell_\Lambda^2} \right)\,.
\end{eqnarray}
We expand $\cg_{\mu\nu}^{(\pm)}$ as
\begin{eqnarray}
 \cg_{\mu\nu}^{(\pm)} &=& \bar{\cg}_{\mu\nu} + 
 \overline{\partial_y \cg_{\mu\nu}} \tilde{y}^{\pm}
 + {1\over2} \overline{\partial^2_y \cg_{\mu\nu}} (\tilde{y}^{\pm})^2 +
 \cdots 
\nonumber \\
 &=& \bar{\cg}_{\mu\nu} -2 \bar{\cK}_{\mu\nu} \tilde{y}^{\pm}
 - \overline{\partial_{\tilde{y}} \cK_{\mu\nu}} (\tilde{y}^{\pm})^2 + \cdots\,,
\label{gmunuexpand}
\end{eqnarray}
where the overbar denotes the value evaluated at $y=0$. 

In the following, we only keep the lowest order in the gradient expansion, 
{\em i.e.}, the leading order of 
\begin{eqnarray}
 \epsilon:=\left|Ky_0 \right| \, ,
\end{eqnarray}
and calculate the effective action 
up to the quadratic order of the metric perturbation.
We want to investigate the parameter region with 
$1\mp 2 r_c^{(\pm)}{\cal H}_\pm>0 $, 
in which both radion and the massive gravitons are healthy~\cite{Yamashita:2014cra}. 
Under the gradient expansion $\epsilon \ll1$, namely, 
$|K|$ becomes $\mathcal{O}\left( 1/ r_c^{(\pm)} \right)$ at most. 
We are also interested in the regime 
in which the mass squared of the lowest mode of massive gravitons 
is comparable to the 4-dimensional momentum squared.
In Ref.~\cite{Yamashita:2014cra}, 
the mass squared of the lowest mode of massive gravitons 
is evaluated as 
\begin{eqnarray}
 \label{masssquared}
 m^2 \simeq {1\over 2r_c^{(+)} y_0}\left( 1 + {1\over \chi} \right) 
 \sim \mathcal{O}\left( {|K|^2/ \epsilon}\right)\,,
\end{eqnarray}
at the lowest order of the gradient expansion. 
Therefore, $H^2$ and the 4-dimensional momentum squared 
become $\mathcal{O}\left( {|K|^2/ \epsilon}\right)$, 
and then 
$\ell_{\Lambda}^{-2}$ should be tuned so that 
$\tilde{\ell}_\Lambda^{-2}
=-\ell_{\Lambda}^{-2} + H^2\sim \mathcal{O}\left( K^2 \right)$ is much
smaller than 
$H^2\sim \mathcal{O}\left( |K|^2/\epsilon \right)$.
This requires the tuning of the background expansion rate $H$ 
to $\approx \ell_\Lambda^{-1}$, 
the value determined by the bulk cosmological constant. 
Furthermore, the Hamiltonian constraint \eqref{HconstEq} requires that 
$R-12 \ell_\Lambda^{-2} \sim \mathcal{O}\left( K^2 \right)$. 
Also the junction conditions \eqref{junction} will imply that 
the traceless part of $R_{\mu\nu}$ becomes 
$\mathcal{O}\left( K^2\right)$, 
while the trace part of $R_{\mu\nu}^{(\pm)}$ cancels with 
the brane tension $\sigma_\pm$
at the leading order of the gradient expansion in Eq.~\eqref{junction}. 
Therefore $R_{\mu\nu}$ is given as 
\begin{eqnarray}\label{Rmunu}
 R_{\mu\nu} \sim 3 \ell_\Lambda^{-2}g_{\mu\nu}+\mathcal{O}\left( K^2 \right)\,.
\end{eqnarray}
From these conditions, the amplitudes of the matter energy momentum 
tensors on the respective branes are severely restricted, which 
does not allow us to use the present model to describe the system 
in which the background energy scale varies by a large amount. 

With this understanding, 
we evaluate $\mathcal{K}_{\mu\nu}$ at the middle point $y=0$ 
from Eq.~\eqref{gmunuexpand} as
\begin{eqnarray}\label{calK}
 \bar{\cK}_{\mu\nu} &=&  -{\cg_{\mu\nu}^{(+)} - \cg_{\mu\nu}^{(-)}\over 4\tilde{y}^+}
 +\cdots \nonumber \\
 &=& \left( {x \over \tilde{\ell}_\Lambda} \gamma_{\mu\nu}
 + {1\over 4y_0}\Delta h_{\mu\nu} \right) (1-x)   
 + \mathcal{O}\left(\epsilon hK \right)\,,
\end{eqnarray}
assuming that $x\sim \mathcal{O}\left( \epsilon^0 h \right)$, 
which will be confirmed later.
Here we define 
\begin{equation}
 \Delta h_{\mu\nu} := h_{\mu\nu}^{(+)}- h_{\mu\nu}^{(-)}\sim
  \mathcal{O}\left( \epsilon h \right)\,.
\label{Deltahmunu}
\end{equation} 
The first term in the parentheses in Eq.~\eqref{calK} comes from 
the factor $a^{-2} (y)$ in the definition of $\cg_{\mu\nu}$~\eqref{cgdef}. 
In order to raise the subscript of $\bar{\cK}_{\mu\nu}$ as well as 
to compute $\bar{R}$, 
we also need $\mathcal{G}_{\mu\nu}$ evaluated at the middle point, which is also 
computed from \eqref{gmunuexpand} 
as 
\begin{eqnarray}\label{bartilg}
 \bar{\cg}_{\mu\nu} &=& {1\over2} \left( \cg_{\mu\nu}^{(+)} + \cg_{\mu\nu}^{(-)} \right) 
 + \overline{\partial_{\tilde{y}} \cK_{\mu\nu}} (\tilde{y}^+)^2 +\cdots \nonumber \\
 &=&
  \gamma_{\mu\nu} + \tilde{h}_{\mu\nu}
  + 2 H^2 x y_0^2 \gamma_{\mu\nu}
  + H^2  \left(x y_0\right)^2\gamma_{\mu\nu} \nonumber \\
  && \qquad \qquad \qquad
  + (1+x)\left( \nabla_\mu \nabla_\nu x \right) y_0^2 
  +\mathcal{O}\left( \epsilon^2,\, \epsilon h \right) \,,
\end{eqnarray}
where $\nabla_\mu$ is the covariant differentiation associated with $\gamma_{\mu\nu}$ 
and we define 
\begin{eqnarray}
 \tilde{h}_{\mu\nu} &:=& {1\over2} \left(h_{\mu\nu}^{(+)} +
				    h_{\mu\nu}^{(-)} \right)\,. 
\label{tildehmunu}
\end{eqnarray}
Here we also use Eq.~\eqref{tKevolEq} but only its last three terms 
where $\ell_\Lambda^{-2}$  and $\bar{R}_{\mu\nu}$ are replaced by 
$H^2$ and $3H^2 \bar{g}_{\mu\nu}$, respectively, 
contribute to \eqref{bartilg} at the order of the present approximation. 
At the 1st order of the metric perturbation $h$, 
the Hamiltonian constraint \eqref{tHconstEq} determines $x^{(1)}$ as 
\begin{eqnarray}\label{x1}
 x^{(1)}  =-{\tilde{\ell}_{\Lambda} \over 16y_0}\Delta h
 - {\tilde{\ell}_{\Lambda}^2 \over 24}\bar{R}^{(1)} +\mathcal{O}\left(  \epsilon h \right)
 =: -{\tilde{\ell}_{\Lambda} \over 16y_0}\Phi +\mathcal{O}\left( \epsilon h \right) \,, 
\end{eqnarray}
where $x$ and $\bar{R}$ are expanded as $A=A^{(1)} + A^{(2)}+\cdots$ 
with respect to order of the metric perturbation $h$.
The last equality defines a new variable $\Phi$, which represents
the rescaled perturbation of the brane separation. 
$\bar{R}^{(1)}$ is obtained from Eq.~\eqref{bartilg} as
\begin{eqnarray}
 \bar{R}^{(1)} &=& \mathcal{L}^{\mu\nu} \left( \tilde{h}_{\mu\nu} 
  + 2 H^2 y_0^2 x^{(1)} \gamma_{\mu\nu} \right) 
  +\mathcal{O}\left(\epsilon h K^2 \right) \nonumber \\
  &=& {1\over 1-\hat{H}^2 \hat{\Box}}\mathcal{L}^{\mu\nu} \left( \tilde{h}_{\mu\nu} 
  - {\hat{H}^2\over 4} \Delta h \gamma_{\mu\nu} \right)
  +\mathcal{O}\left(\epsilon h K^2 \right)\,,
\label{barR1}
\end{eqnarray}
where we define 
\begin{eqnarray}
\mathcal{L}_{\mu\nu} := \nabla_\mu \nabla_\nu - {\Box \over4} \gamma_{\mu\nu} 
 - {3\over4} \left( \Box+4H^2 \right)\gamma_{\mu\nu}\,,  \qquad  
 \Box := \nabla^\mu \nabla_\mu\,,
 \end{eqnarray}
and 
\begin{eqnarray*}
&& \alpha := {y_0 \tilde{\ell}_\Lambda \over 2}\,,\qquad
 \hat{H}^2 := \alpha H^2\,,\qquad
\hat{\Box} := \alpha \left( \Box+4H^2\right)\,.
\end{eqnarray*}
To arrive at the last expression in Eq.~\eqref{barR1}, 
we formally solved the 
differential equation obtained by substituting Eq.~\eqref{x1} 
into the first equality in Eq.~\eqref{barR1}, and 
we also use the identity 
\begin{equation}
 \mathcal{L}_{\mu\nu}\nabla^\mu A^\nu= 0\,,
\label{usedidentity}
\end{equation}
which holds for an arbitrary vector $A^{\nu}$.
Here, we need to discuss the 
boundary conditions for the inverse of the differential operator 
$1-\hat{H}^2 \hat{\Box}$. 
One may think that 
the non-local operator $(1-\hat{H}^2\hat{\Box})^{-1}$ may bring an
extra degree of freedom corresponding to the pole 
at $\hat{\Box}=\hat{H}^{-2}$. 
However, we know that there is no physical mode at
$\hat{\Box}=\hat{H}^{-2}$ from the linear perturbation 
analysis~\cite{Izumi, Yamashita:2014cra}.  
In order to avoid the appearance of such an unphysical degree of freedom, 
we should understand that the non-local operator 
$(1-\hat{H}^2\hat{\Box})^{-1}$  
is to be expanded as $1+\hat{H}^2 \hat{\Box}+ \hat{H}^4 \hat{\Box}^2+\cdots$ 
by restricting the energy scale to $\hat{H}^2 \hat{\Box} < 1$. 
When $H \approx m$ and the normal branch is chosen: 
$1\mp 2 r_c^{(\pm)}{\cal H}_\pm>0 $, 
the allowed energy region where our approximation is justified 
is estimated as 
\begin{eqnarray}\label{rest}
 \Box+4H^2 \lesssim \mathcal{O}\left( K^2 {r_c^{(+)} \over y_0} \right) 
 \lesssim \mathcal{O}\left( m^2 \right)\,,
\end{eqnarray}
which is marginally consistent with the energy scale we wish to
discuss. Interestingly, we will find later that, even if we keep 
this pole at $\hat{\Box}=\hat{H}^{-2}$, its contribution to 
the metric perturbation is not sourced by the matter energy momentum 
tensors localized on the branes. 
Substituting Eq.~\eqref{barR1} into Eq.~\eqref{x1}, 
we obtain
\begin{eqnarray}\label{Phi1}
 \Phi = -{16y_0 \over \tilde{\ell}_\Lambda} x^{(1)} 
 = \Delta h + {4\alpha \over3} {1\over 1-\hat{H}^2 \hat{\Box}}\mathcal{L}^{\mu\nu}
  \left( \tilde{h}_{\mu\nu}  - {\hat{H}^2\over 4} \Delta h \gamma_{\mu\nu} \right)\,. 
\end{eqnarray}

Equipped with the solution to the linear order in $\epsilon$, 
we can evaluate the Hamiltonian constraint \eqref{tHconstEq} at the 2nd order of $h$ as 
\begin{eqnarray}
  &-& {6\over \tilde{\ell}_{\Lambda}} \left[ {4x^{(2)} \over \tilde{\ell}_{\Lambda}}
  - \left({\Delta h\over 4y_0} + {4x^{(1)} \over \tilde{\ell}_{\Lambda}} \right)x^{(1)} \right] 
  \nonumber \\
 && \qquad\qquad - {12x^{(1)\,2} \over \tilde{\ell}_{\Lambda}^2} 
 - {x^{(1)} \Delta h \over 2\tilde{\ell}_{\Lambda}y_0} 
 -  {\Delta h^{\mu\nu} \Delta h_{\mu\nu} \over 16 y_0^2 } 
 + \mathcal{O}\left(\epsilon h^2 K^2 \right)
 = \bar{R}^{(2)}\,,
\end{eqnarray}
where $\bar{R}^{(2)}$ can be more explicitly expressed 
using Eq.~\eqref{bartilg} as
\begin{eqnarray}\label{R2}
 \bar{R}^{(2)} = y_0^2\mathcal{L}^{\mu\nu}
 \left[ x^{(1)} \nabla_\mu \nabla_\nu x^{(1)} + H^2 \left(x^{(1)}\right)^2 \gamma_{\mu\nu}
 +2H^2 x^{(2)}\gamma_{\mu\nu} \right] + \mathcal{O}\left(\epsilon h^2 K^2 \right)\,.
 \end{eqnarray}
Then, as in the case of $x^{(1)}$, $x^{(2)}$ 
is solved as 
\begin{eqnarray}
x^{(2)} &=& {1\over 1-\hat{H}^2 \hat{\Box}}
 {\tilde{\ell}_{\Lambda}^2 \over 384 y_0^2} 
 \left[ \Delta h^2- \Delta h^{\mu\nu} \Delta h_{\mu\nu} - {3\over4}\Phi^2 
 +4\alpha \Phi \bar{R}^{(1)} \right.  \nonumber \\
 &&\qquad\qquad\qquad\qquad\qquad\qquad
 \left. -{\alpha^2 \over 4} \mathcal{L}^{\mu\nu} 
 \left( \Phi \nabla_\mu \nabla_\nu \Phi + H^2 \Phi^2 \gamma_{\mu\nu} \right) \right]
 + \mathcal{O}\left(\epsilon h^2 \right)\,.
\end{eqnarray}
Therefore, the total second-order effective action 
\begin{eqnarray}
\label{noaction}
 S  &=& {M_{pl}^2\over2}\Biggl[{1 \over 2r_c^{(+)}} \oint d^5x 2\sqrt{-g}
			 \,  \left( R- { 12 \over \ell_{\Lambda}^2 }
			     \right)  \cr
&&\qquad
+ \int d^4x \sqrt{-g_{(+)}} \left( R_{(+)} -{6H^2\over a_+^2} \right) 
 + \chi \int d^4x \sqrt{-g_{(-)}} \left( R_{(-)} - {6H^2 \over a_-^2}
				  \right) \Biggr]
\,,
\end{eqnarray}
is expressed in terms of $g^{(\pm)}_{\mu\nu}$ as 
\begin{eqnarray}
{1 \over 2r_c^{(+)}}\oint & d^5x& 2\sqrt{-g}
			 \,  \left( R- { 12 \over \ell_{\Lambda}^2 }
			     \right)\cr
 &=& 
{1 \over r_c^{(+)}} \int^{-y_0^+}_{y_0^+} Ndy \int d^4x
 2\sqrt{-\gamma}\, \left\{  { 12 \over \tilde{\ell}_{\Lambda}^2} 
 + \mathcal{L}^{\mu\nu} \left( \bar{g}_{\mu\nu} 
 - \tilde{y}^2 \overline{\partial_{\tilde{y}} K_{\mu\nu}}\right) 
  + \mathcal{O}\left( \epsilon h^2 K^2 \right) \right\}\nonumber \\
   &=& 
{1 \over r_c^{(+)}} \int d^4x\,
 2\sqrt{-\gamma}\, \left\{  
 2N y_0 \left( {12\over \tilde{\ell}^2_\Lambda} 
 + \bar{R}^{(1)}+\bar{R}^{(2)} \right)   \right. \nonumber \\
 &&\left.  \qquad\qquad\qquad\qquad\qquad
 - {2\over3}N y_0^3 \mathcal{L}^{\mu\nu} 
 \left( N\nabla_\mu \nabla_\nu N+N^2H^2\gamma_{\mu\nu}\right)
  + \mathcal{O}\left( \epsilon^2 h^2 K \right) \right\}\nonumber \\
 &=& \int d^4x \sqrt{-\gamma}\,
 {1\over 8r_c^{(+)} y_0 } 
 \left( 384 \left({y_0\over \ell_{\Lambda}}\right)^2 \left( 1-4\hat{H}^4 \right) x^{(2)} 
 - 4\alpha \Phi \bar{R}^{(1)} + \mathcal{O}\left( \epsilon^3 h^2 \right) \right)
  \nonumber \\
 &=&\int d^4x \sqrt{-\gamma}\, m_*^2 \left\{
  \Delta h^2 - \Delta h^{\mu\nu} \Delta h_{\mu\nu} 
 - {3\over4}\Phi \left( 1-\hat{H^2}\hat{\Box} \right)\Phi 
 + \mathcal{O}\left( \epsilon^3 h^2 \right) \right\} \,,
\label{effectiveaction}
\end{eqnarray}
where we define 
\begin{equation}
 m_*^2 := {1\over 8 r_c^{(+)} y_0 }\,.
\end{equation}
In the first equality in Eq.~\eqref{effectiveaction}, 
$R(y)- 12 \ell_{\Lambda}^{-2} $ is expanded around $y=0$ as 
\begin{eqnarray}
 R(y)- { 12 \over \ell_{\Lambda}^2} &=& 12H^2  + \mathcal{L}^{\mu\nu} g_{\mu\nu}(y) - { 12 \over \ell_{\Lambda}^2} + \mathcal{O}\left( \epsilon K^2 \right) \nonumber \\
 &=& { 12 \over \tilde{\ell}_{\Lambda}^2} +\mathcal{L}^{\mu\nu}\left(\bar{g}_{\mu\nu} -2\tilde{y}\bar{K}_{\mu\nu} -\tilde{y}^2 \overline{\partial_{\tilde{y}} K_{\mu\nu} } +\cdots\right) + \mathcal{O}\left( \epsilon K^2 \right) \,,
\end{eqnarray}
where the linear term in $\tilde{y}$ is integrated to be zero in the integral. 
In the third equality in Eq.~\eqref{effectiveaction}, 
we use Eqs.~\eqref{usedidentity} and \eqref{R2} 
and neglect the total derivative terms. 
The action \eqref{noaction} contains $(1-\hat{H}^2\hat{\Box}
)^{-1}$ through $\Phi$ defined by Eq.~\eqref{Phi1}. 
As mentioned above, we treat the apparently non-local operator 
$( 1-\hat{H}^2\hat{\Box})^{-1}$ as a 
local one by imposing the restriction presented in Eq.~\eqref{rest}. 

In the extreme limit of self-accelerating branch $\alpha m_*^2 \rightarrow 0$, 
which corresponds to the limit $r_c^{(+)}\mathcal{H}_+ \rightarrow \infty$, 
$\Phi$ is reduced to $\Delta h$ and the bulk contribution in the 
effective action \eqref{effectiveaction} becomes 
\begin{equation}
 {M_{pl}^2 m_*^2 \over2}\int d^4x \sqrt{-\gamma}\, 
\left( {1\over4}\Delta h^2-\Delta h_{\mu\nu}\Delta h^{\mu\nu} \right)\,,
\nonumber
\end{equation} 
which differs from the form of the ghost-free dRGT mass terms. 
Therefore, the system possesses an extra scalar mode in addition to the
two gravitons and a BD ghost appears. 
In the self-accelerating branch of DGP 2-brane model, 
we know that either the extra scalar mode 
corresponding to the radion or the helicity-0 mode of the massive 
graviton becomes ghost~\cite{Charmousis, Izumi}. 
We refer to them as the radion ghost and the Higuchi ghost, respectively. 
The appearance of a BD ghost here seems to be exactly corresponding to 
the inevitable existence of either the radion ghost or the Higuchi
ghost.  
When the mass squared of the massive graviton is smaller than $2H^2$, 
the ghost in the DGP model is the Higuchi ghost, and the parameter 
region with such a small graviton mass is not excluded. 
Hence, this example shows that the BD ghost 
can be the Higuchi ghost. The existence of the Higuchi ghost 
may not be so harmful as a usual scalar ghost, as discussed 
in Ref.~\cite{Izumi2}.

Now we return to the case without taking 
the limit $\alpha m_*^2 \rightarrow 0$, 
and show in general that the system described by the total
effective action~\eqref{noaction} consists of 
one massless graviton, one massive graviton, and one scalar.
We rewrite the action \eqref{noaction} 
so that the rescaled perturbation of the brane separation, 
$\Phi$, 
is manifestly treated as an independent degree of freedom, radion. 
For this purpose, we introduce a Lagrange multiplier $\lambda$ to 
impose the constraint $\Phi = \Delta h + {4\alpha \over3}\bar{R}^{(1)}$ as
\begin{eqnarray}\label{actionwithL}
 S &=& {M_{pl}^2\over 2} \left[ \int d^4x \sqrt{-\gamma}\, m_*^2 \left\{
  \Delta h^2 - \Delta h^{\mu\nu} \Delta h_{\mu\nu} 
 - {3\over4}\Phi \left( 1-\hat{H^2}\hat{\Box} \right)\Phi 
 + \lambda \left( \Phi - \Delta h - {4\alpha \over3}\bar{R}^{(1)} \right) \right\} \right. \nonumber \\
 &&\left. \qquad\qquad+ \int d^4x \sqrt{-g_{(+)}} \left( R_{(+)} -{6H^2\over a_+^2} \right) 
 + \chi \int d^4x \sqrt{-g_{(-)}} \left( R_{(-)} - {6H^2 \over a_-^2} \right) \right]\,.
\end{eqnarray}
Taking the variation with respect to $\Phi$, we obtain a 
simple expression for $\lambda$ written in terms of $\Phi$:
\begin{eqnarray}
 \lambda={3\over2} \left( 1-\hat{H}^2\hat{\Box} \right) \Phi\,.
\label{lambdaeq}
\end{eqnarray}
Then, substituting this $\lambda$ back into the action \eqref{actionwithL}, 
the action written in terms of $h_{\mu\nu}^{(\pm)}$ and $\Phi$ is obtained as 
\begin{eqnarray}\label{actionPhi}
 S &=& {M_{pl}^2\over 2} \left[ \int d^4x \sqrt{-\gamma}\, m_*^2 \left\{
  \Delta h^2 - \Delta h^{\mu\nu} \Delta h_{\mu\nu} 
 - {3\over4}\alpha^2 H^2\Phi \left( \Box+4H^2 \right)\Phi  \right.\right. \nonumber \\
 &&\left.\left. \qquad\qquad\qquad\qquad\qquad\qquad\qquad\qquad
 + {3\over4} \Phi\left(\Phi -2\Delta h \right)
 - 2\alpha \Phi \mathcal{L}^{\mu\nu} \tilde{h}_{\mu\nu}
 \right\}  \right. \nonumber \\
 &&\left. \qquad\,\,
 + \int d^4x \sqrt{-g_{(+)}} \left( R_{(+)} -{6H^2\over a_+^2} \right) 
 + \chi \int d^4x \sqrt{-g_{(-)}} \left( R_{(-)} - {6H^2 \over a_-^2} \right) \right]\,.
\end{eqnarray}
In this step, 
$1-\hat{H}^2\hat{\Box}$ in Eq.~\eqref{lambdaeq} cancels out 
the apparently non-local operator $(1-\hat{H}^2\hat{\Box})^{-1}$ 
in $\bar{R}^{(1)}$. 
After rewriting the effective action into this form, 
we can obtain the ghost-free Fierz-Pauli model 
even in the limit $\alpha m_*^2 \rightarrow 0$, 
by setting the radion $\Phi$ to 0, assuming some steep stabilization 
potential for $\Phi$ by hand. 
The last term in the second line of Eq.~\eqref{actionPhi} can be absorbed 
into the induced gravity terms by the conformal transformations 
$g^{(+)}_{\mu\nu} = \exp{\left(m_*^2\alpha \Phi \right)} \hat{g}_{\mu\nu}^{(+)}$ and 
$g^{(-)}_{\mu\nu} = \exp{\left([{m_*^2\alpha/\chi}] \Phi \right)} \hat{g}_{\mu\nu}^{(-)}$. 
Then, the action becomes
\begin{eqnarray} \label{linearactionf}
 S &=& {M_{pl}^2\over 2} \left[ \int d^4x \sqrt{-\gamma}\, m_*^2 \left\{
  \Delta h^2 - \Delta h^{\mu\nu} \Delta h_{\mu\nu} 
 + {3\over4} \Phi\left(\Phi -2\Delta h \right)   \right. \right. \nonumber \\
 &&\left.\left. \qquad\qquad\qquad\qquad\qquad
 + {3\over2}\alpha^2 \left( m_*^2 \left( 1+ {1\over \chi} \right) - {1\over2}H^2 \right)
 \Phi \left( \Box+4H^2 \right)\Phi
  \right\} \right. \nonumber \\
 &&\left. \qquad\,\,+ \int d^4x \sqrt{-\hat{g}_{(+)}} \left( \hat{R}_{(+)} -{6H^2\over a_+^2} \right) 
 + \chi \int d^4x \sqrt{-\hat{g}_{(-)}} \left( \hat{R}_{(-)} - {6H^2 \over a_-^2} \right) \right]\,,
\end{eqnarray}
where $\hat{R}_{(+)}$ and $\hat{R}_{(-)}$ 
are the Ricci scalars for the metrics 
$\hat{g}^{(+)}_{\mu\nu}$ and $\hat{g}^{(-)}_{\mu\nu}$, respectively.
Here, two gravitons interact with each other through the 
Fierz-Pauli mass term and the radion field $\Phi$, which is now an 
independent field, couples to both metrics, 
but its kinetic term couples to the metrics only through 
the combination $\gamma_{\mu\nu}$. 
To be honest, at the current level of our approximation in which 
all the terms higher order in $\epsilon$ are neglected, we cannot 
discriminate to which metric the radion field is coupled.
Therefore the model described by this effective action 
satisfies the BD-ghost-free conditions for doubly coupled fields in bigravity~\cite{Yamashita:2014fga, deRham:2014naa}. 
Since the action \eqref{linearactionf} does not suffer from higher derivatives nor BD ghost problem, 
this system contains only two gravitons and one scalar radion. 
Because of the presence of the constraints for $\Delta h_{\mu\nu}$, 
it cannot be immediately confirmed by the action \eqref{linearactionf} 
whether the helicity-0 mode of the massive graviton and the scalar radion are healthy or not, 
which will be discussed in the next section 
by writing down the effective action in terms of the matter energy-momentum tensor, 
integrating out the gravitational degrees of freedom.

\section{Equations of motion and coupling to the matter 
energy momentum tensors on the branes}

Now we analyze the equations of motion derived from the action~\eqref{noaction} 
with additional matter fields localized on the branes. 
We confirm that the system contains only two gravitons and one scalar mode 
whose ghost-free condition is equivalent to the one discussed in 
Ref.~\cite{Yamashita:2014cra}, 
by investigating the poles of the propagators and their coefficients. 
Taking the variations of the action \eqref{noaction} with respect to $h_{\mu\nu}^{(\pm)}$, we obtain the equations of motion:
\begin{eqnarray}
\label{EOMpm}
 &&\chi_\pm \mathcal{E}_{\mu\nu}^{\alpha\beta} h_{\alpha\beta}^{(\pm)} 
 \pm 2{m_*^2} \left( \Delta h_{\mu\nu} - \Delta h \gamma_{\mu\nu} \right)  \nonumber \\
 &&\qquad\qquad
 + {3\over2}m_*^2  \left[ \pm\gamma_{\mu\nu} + {2\over3}\alpha \mathcal{L}_{\mu\nu}  \right]
 \left( \Delta h + {4\over3}\alpha \bar{R}^{(1)} \right) = M_{pl}^{-2} T^{(\pm)}_{\mu\nu} \,, 
\end{eqnarray}
where we define 
$\chi_+=1$, $\chi_-=\chi$, and
\begin{eqnarray}
\mathcal{E}_{\mu\nu}^{\alpha\beta} h_{\alpha\beta} &:= &
-{1\over2} \left( \Box h_{\mu\nu} + \nabla_\mu \nabla_\nu h 
- 2\nabla_{(\nu} \nabla_{\sigma} h^\sigma_{\mu)} - \gamma_{\mu\nu}\Box h 
+ \gamma_{\mu\nu} \nabla_\alpha \nabla_\beta h^{\alpha\beta} \right. \nonumber \\
&&\left. \qquad\qquad\qquad\qquad\qquad\qquad\qquad\qquad\qquad
-2H^2h_{\mu\nu} - H^2 \gamma_{\mu\nu} h \right)\,, 
\end{eqnarray}
is the linearized Einstein equations derived from the variation 
of the four-dimensional localized Einstein-Hilbert part of the action. 
In the above expression, we set $a_\pm=1$. 
In the present order of approximation, it is no use to distinguish 
$a_\pm$ from unity.
Here, we impose the gauge conditions 
\begin{equation}
\nabla^\mu \left( \tilde{h}_{\mu\nu} - {1\over4}\gamma_{\mu\nu}
	    \tilde{h} \right)=0\,,
\end{equation}
on the averaged part of the metric perturbation~\eqref{tildehmunu}, 
so that the traceless part of $\tilde{h}_{\mu\nu}$ becomes transverse. 
On the other hand, we decompose the difference of the 
metric perturbations \eqref{Deltahmunu} as 
\begin{eqnarray}
 \Delta h_{\mu\nu} = \Delta h_{\mu\nu}^{(TT)} + {1\over4} \phi \gamma_{\mu\nu}
 + 
\left( \nabla_\mu \nabla_\nu - {\Box \over4} \gamma_{\mu\nu} \right)
 \psi
\,, 
\label{Deltah}
\end{eqnarray}
where $\Delta h_{\mu\nu}^{(TT)}$ is the transverse-traceless part of 
$\Delta h_{\mu\nu}$. 
Using the identities, \eqref{usedidentity} and 
\begin{eqnarray}
&&\nabla^\mu \mathcal{E}_{\mu\nu}^{\alpha\beta} H_{\alpha\beta}=0\,,
\\
\label{idscalar}
&&\nabla^\mu 
\left( \nabla_\mu \nabla_\nu - {\Box \over4} \gamma_{\mu\nu} \right)
 \Psi
={3\over 4}\nabla_\nu \left(\Box+4H^2 \right)\Psi\,,\\
&& \nabla^\mu \mathcal{L}_{\mu\nu} \Psi=0\,,
\end{eqnarray}
where $\Psi$ and $H_{\alpha\beta}$ are arbitrary 
scalar and tensor, respectively, 
one can take the divergence of Eq.~\eqref{EOMpm}.  
Then, the two equations derived from the divergence of 
Eq.~\eqref{EOMpm} give the same equation, 
from which $\psi$ is determined as
\begin{eqnarray}\label{div}
 \psi = {\alpha \over2} Z \, ,
\end{eqnarray}
where 
\begin{eqnarray}
 \label{Z}
 Z := \left( 1- \hat{H}^2 \hat{\Box} \right)^{-1} \left( 2\tilde{h} -2\hat{H}^2 \phi \right) \,.
\end{eqnarray}
Here, we ignore the homogeneous solution that satisfies $\left(
\Box+4H^2 \right)\psi=0$ 
because it degenerates with the transverse-traceless
mode $\Delta h_{\mu\nu}^{(TT)}$ by the identity \eqref{idscalar}. 
One may think that $\psi$ is not uniquely determined 
because of the pole at $\hat{\Box} =\hat H^{-2}$.
However, such a possible ambiguity has been already eliminated by 
imposing the restriction~\eqref{rest}. 
By substituting Eq.~\eqref{div}, Eq.~\eqref{EOMpm} gives
\begin{eqnarray}\label{cEOMpm}
 \chi_\pm \mathcal{E}_{\mu\nu}^{\alpha\beta} h_{\alpha\beta}^{(\pm)} 
 \pm 2m_*^2 \Delta h_{\mu\nu}^{(TT)}
 + m_*^2 \alpha \mathcal{L}_{\mu\nu} 
 \left[ \phi \pm\left( 1\mp{\hat{\Box} \over2} \right)Z \right] 
 = M_{pl}^{-2} T^{(\pm)}_{\mu\nu} \,.
\end{eqnarray}
On the other hand, 
the trace of the metric perturbation on each brane $h^{(\pm)}$ 
after eliminating the terms proportional to $\nabla_\mu \nabla_\nu
\psi$, 
which can be erased by a 4-dimensional gauge transformation from $h_{\mu\nu}^{(\pm)}$, 
are found to be given by
\begin{eqnarray} \label{trhpm}
 2h^{(\pm)} &:=& 2\tilde{h} \pm \left( \phi - \Box \psi \right) \nonumber \\
 &=& \left( 2\hat{H}^2 \pm 1 \right) 
 \left[\phi \pm\left( 1\mp{\hat{\Box} \over2} \right)Z\right]\,,
\end{eqnarray} 
where in the second equality we use Eq.~\eqref{Z} again. 
Using this equality and Eq.~\eqref{Hpmexpand}, 
the equations of motion for $h^{(\pm)}_{\mu\nu}$ \eqref{cEOMpm} 
are simplified as 
\begin{eqnarray}\label{EOM}
 \chi_\pm \mathcal{E}_{\mu\nu}^{\alpha\beta} h_{\alpha\beta}^{(\pm)} 
 \pm 2m_*^2 \Delta h_{\mu\nu}^{(TT)}
 + \left( \pm8r_c^{(+)}\mathcal{H}_\pm \right)^{-1} \mathcal{L}_{\mu\nu} h^{(\pm)}
 = M_{pl}^{-2} T^{(\pm)}_{\mu\nu} \,. 
\end{eqnarray}
The trace part of Eq.~\eqref{EOM} becomes
\begin{eqnarray}
 h^{(\pm)} = {4M_{pl}^{-2}\over3} 
 \left(\chi_\pm \mp {1\over 2r_c^{(+)}\mathcal{H}_\pm} \right)^{-1} 
 {1\over \Box+4H^2} T^{(\pm)}\,.
\end{eqnarray}
Diagonalizing the traceless part of Eq.~\eqref{EOM} with respect to the mass eigenvalues, 
we obtain the equations of motion for the transverse-traceless part of 
the massless mode $h_{\mu\nu}^{(0)}$, 
\begin{equation}
 h_{\mu\nu}^{(0)} := \left( 1+ \chi \right)^{-1} \left(
		     h_{\mu\nu}^{(+)} + \chi h_{\mu\nu}^{(-)}
							  \right)\,,
\end{equation}
and that of the massive mode $h_{\mu\nu}^{(m)}$
\begin{equation}
 h_{\mu\nu}^{(m)} := \left(1 + \chi \right)^{-1} \Delta
  h_{\mu\nu}\,,
\end{equation}
 as
\begin{eqnarray}
 \left(\Box - 2H^2\right) h_{\mu\nu}^{(0)\,TT} &=& {-2M_{pl}^{-2} \over 1 + \chi }
 \left[T_{\mu\nu}^{(0)} - {1\over4} T^{(0)}\gamma_{\mu\nu} \right. \nonumber \\
 && \left. \qquad\qquad\quad
  + {1\over3} \left( \nabla_\mu \nabla_\nu - {\Box\over4}\gamma_{\mu\nu}\right)
   {1\over \Box+4H^2}T^{(0)} \right]\,,
  \\
  \left(\Box - 2H^2 -m^2 \right) h_{\mu\nu}^{(m)\,TT} &=& {-2M_{pl}^{-2}
   \over 1 + \chi } 
  \left[T_{\mu\nu}^{(m)} - {1\over4} T^{(m)}\gamma_{\mu\nu} \right. \nonumber \\
 && \left.  \qquad\qquad\quad
  + {1\over3} \left( \nabla_\mu \nabla_\nu - {\Box\over4}\gamma_{\mu\nu}\right)
   {1\over \Box+4H^2}T^{(m)} \right]\,,
\end{eqnarray}
where $m^2$ is defined in Eq.~\eqref{masssquared} and we define
\begin{eqnarray}
 T^{(0)}_{\mu\nu}&:=& T^{(+)}_{\mu\nu}+T^{(-)}_{\mu\nu}\,, \\ 
 T^{(m)}_{\mu\nu}&:=& T^{(+)}_{\mu\nu} - {T^{(-)}_{\mu\nu}\over \chi}\,, 
 \end{eqnarray}
Using the identity
\begin{eqnarray}
 \left(\nabla_\mu \nabla_\nu - {\Box\over4}\gamma_{\mu\nu}\right){1\over \Box+4H^2}\Psi
 ={1\over \Box-4H^2}\left( \nabla_\mu \nabla_\nu - {\Box\over4}\gamma_{\mu\nu} \right)\Psi \,,
\end{eqnarray}
which holds for an arbitrary scalar $\Psi$,
$h_{\mu\nu}^{(0)}$ and $h_{\mu\nu}^{(m)}$ are found to be rewritten as 
\begin{eqnarray}
 h_{\mu\nu}^{(0)\,TT} &=& {-2M_{pl}^{-2} \over 1 + \chi }
 \left[{1\over \Box - 2H^2}\left\{ T_{\mu\nu}^{(0)} - {1\over4} T^{(0)}\gamma_{\mu\nu} 
  + {1\over3\left(-2H^2\right)} \left( \nabla_\mu \nabla_\nu - {\Box\over4}\gamma_{\mu\nu}\right)T^{(0)} \right\} \right. \nonumber \\
  &&\left. \qquad\qquad\qquad
  -{1\over3\left(-2H^2\right)}\left( - {\Box\over4}\gamma_{\mu\nu}\right) {1\over \Box+4H^2}T^{(0)} \right]\,,
  \\
 h_{\mu\nu}^{(m)\,TT} &=& {-2M_{pl}^{-2} \over 1 + \chi } 
  \left[{1\over \Box - 2H^2-m^2}\left\{ T_{\mu\nu}^{(m)} - {1\over4} T^{(m)}\gamma_{\mu\nu} 
  \right. \right. \nonumber \\
  && \left. \left. \qquad\qquad\qquad\qquad\qquad\qquad
  + {1\over3\left(m^2-2H^2\right)} \left( \nabla_\mu \nabla_\nu - {\Box\over4}\gamma_{\mu\nu}\right)T^{(m)} \right\} \right. \nonumber \\
  &&\left. \qquad\qquad\qquad
  -{1\over3\left(m^2-2H^2\right)}\left( - {\Box\over4}\gamma_{\mu\nu}\right){1\over \Box+4H^2}T^{(m)} \right]\,,
\end{eqnarray}
where we eliminate some terms 
which can be erased by a 4-dim gauge transformation as in Eq.~\eqref{trhpm}. 
If we set $T^{(-)}_{\mu\nu}=0$, for simplicity, 
the coefficient of the pole ${(\Box+4H^2)^{-1}}$ in $h_{\mu\nu}^{(+)}
=h_{\mu\nu}^{(0)}+\chi h_{\mu\nu}^{(m)}$
is extracted as 
\begin{eqnarray}\label{sourcedh+}
  h_{\mu\nu}^{(+)}\supset {M_{pl}^{-2}\over3}
  \left[ {2r_c^{(+)}\mathcal{H}_+ \over 2r_c^{(+)}\mathcal{H}_+ -1} + 
 {1\over 1 + \chi} \left( -1+
{2\chi H^2\over m^2-2H^2} \right)\right] 
 {1\over \Box+4H^2}T^{(+)}\gamma_{\mu\nu} \,. 
\end{eqnarray}
When we integrate out the scalar degree of freedom, 
the contribution of the pole at $\Box=-4H^2$ 
to the effective action is written in terms of $T_{\mu\nu}^{(+)}$ as 
\begin{eqnarray}
 -\int d^4x \sqrt{-\gamma}\, h_{\mu\nu}^{(+)}T^{\mu\nu}_{(+)} \supset 
  {M_{pl}^{-2}\over3} \beta_+ \int d^4x \sqrt{-\gamma}\, T^{(+)}{1\over
  \Box+4H^2}T^{(+)} \,,
\end{eqnarray}
where 
\begin{eqnarray}
 \beta_+
 &:=&  
{1\over 
1-2r_c^{(+)}\mathcal{H}_+} -
 {\chi \over 1 + \chi} {m^2\over m^2-2H^2}\,,
\label{beta+}
\end{eqnarray}
is the minus of the part embraced by 
the square brackets in Eq.~\eqref{sourcedh+}. 
The sign of $\beta_+$ determines whether 
the scalar degree of freedom is ghost or not. 
When $\beta_+$ is negative, the scalar degree of freedom is ghost. 
When the condition $1-2r_c^{(+)}\mathcal{H}_+ > 0$,
which is one of the two ghost-free branch conditions 
derived in Ref.~\cite{Yamashita:2014cra}, is satisfied, 
the first term in the square brackets in Eq.~\eqref{beta+} is 
positive and greater than unity. %
If ${\cal H}_+$ is positive, for sufficiently large $m^2$, the absolute value 
of the second term becomes less than unity, and hence 
$\beta_+$ becomes positive in total. 
When $1-2r_c^{(+)}\mathcal{H}_+ < 0$, the positiveness of $\beta_+$ 
imposes $m^2-2H^2<0$, namely, Higuchi ghost appears. 
 
A similar discussion applies to the case where $T^{(+)}_{\mu\nu}=0$. 
The coefficient of the pole $({\Box+4H^2})^{-1}$ in $h_{\mu\nu}^{(-)}$ becomes
\begin{eqnarray} \label{sourcedh-}
 - \int d^4x \sqrt{-\gamma}\, h_{\mu\nu}^{(-)}T^{\mu\nu}_{(-)} \supset 
  {M_{pl}^{-2}\over3} \beta_- \int d^4x \sqrt{-\gamma}\, T^{(-)}{1\over
  \Box+4H^2}T^{(-)} \,,
\end{eqnarray}
where
\begin{eqnarray}
 \beta_- := {1\over \chi}
\left[
  {1 \over 1+2r_c^{(-)}\mathcal{H}_- } -
 {1\over 1 + \chi} {m^2\over m^2-2H^2}\right] \,.
\label{beta-}
\end{eqnarray}
Thus, the contribution of the pole at $\Box=-4H^2$ 
to the effective action 
is positive and free from Higuchi ghost 
when $1+2r_c^{(-)}\mathcal{H}_- >0$ 
and the first term is larger than the second term in magnitude.  
The first condition is identical to the other ghost-free-branch condition 
derived in Ref.~\cite{Yamashita:2014cra}, as is expected. 

Suppose that for sufficiently large $m^2$, the ghost-free conditions are
satisfied. 
If we set $m^2$ is larger than and close to $2H^2$, 
both $\beta_+$ and $\beta_-$ 
necessarily become negative. Hence, 
there should be critical values of $m^2$ at which $\beta_+$ or $\beta_-$ 
cross zero. Since the scalar mode couples to both traces of the 
energy momentum tensors, $T_{(+)}$ and $T_{(-)}$, we expect that 
the signatures of $\beta_+$ and $\beta_-$ should flip simultaneously. 
In fact, at the leading order of the gradient expansion, 
we can rewrite $\beta_\pm$ 
using Eqs.~\eqref{Hpmexpand} and \eqref{masssquared} as
\begin{eqnarray}
 \beta_\pm = {1\over 1+\chi} {m^2\over m^2 -2H^2} 
 {1\pm 2r_c^{(\mp)}\mathcal{H}_\mp \over 1\mp 2r_c^{(\pm)}\mathcal{H}_\pm}\,.
\end{eqnarray}
Then, we find that
the condition $\beta_\pm>0 $ is equivalent to 
$1\pm 2r_c^{(\mp)}\mathcal{H}_\mp>0$, when $m^2>2H^2$ 
and $1\mp 2r_c^{(\pm)}\mathcal{H}_\pm>0$. 
This simultaneous sign flip occurs at the transition from the ghost-free 
normal branch to the self-accelerating branch. 
At the transition point in the parameter space, either $\beta_+$ 
or $\beta_-$ diverges. (If we assume ${\cal H}_+\approx {\cal H}_->0$, 
$\beta_-$ never diverges. Then, the divergent is necessarily $\beta_+$.) 
Therefore, the transition between two branches is accompanied by
the divergence of metric perturbation and 
the perturbative approach breaks down at the branch crossing point. 

Let us examine how one can construct 
the background spacetime that satisfies 
the conditions for the normal branch. 
Using the normal-branch conditions $1\mp2r_c^{(\pm)}\mathcal{H}_\pm \geq 0$ 
and Eqs.~\eqref{Hpmexpand} and \eqref{masssquared}, 
we obtain a constraint for $\tilde{\ell}_\Lambda$ as 
\begin{eqnarray}\label{normalcond}
 -1+\left( 1+\chi \right){H^2 \over m^2} < {2r_c^{(-)} \over \tilde{\ell}_\Lambda}
 < \chi-\left( 1+\chi \right){H^2 \over m^2}\,.
\end{eqnarray}
In order to satisfy this constraint,  
$-1+\left( 1+\chi \right)H^2 / m^2 < \chi-\left( 1+\chi \right) H^2 / m^2$ must be satisfied. 
Thus, the mass squared of the massive graviton \eqref{masssquared}, 
which can be adjusted by changing $y_0^+$, 
must be tuned to be above the Higuchi bound, {\em i.e.}, $m^2>2H^2$. 
When $m^2>2H^2$, Eq.~\eqref{normalcond} can be achieved 
by tuning $\tilde{\ell}_\Lambda$ depending on $H/m$ and $r_c^{(\pm)}$. 

The physical meaning of the normal branch conditions
can be made clear in the following way. 
Eliminating $H^2/a_\pm^2$ from Eq.~\eqref{jcge1} with \eqref{Hge1}, 
we obtain the relations among 
$\mathcal{H}_\pm$, $\sigma_\pm$ and $\ell_\Lambda$ as 
\begin{eqnarray}
 \mathcal{H}_\pm^2 \mp {\mathcal{H}_\pm \over r_c^{(\pm)}} 
 + {1\over \ell_\Lambda^2} - {\sigma_\pm \over 3} =0\,. 
\label{junctioncomp}
\end{eqnarray}
Then, we find that there are two solutions for $\mathcal{H}_\pm$ as
\begin{eqnarray}\label{branches+}
 1-2r_c^{(+)}\mathcal{H}_+ = \pm \sqrt{1-4r_c^{(+)\,2} 
 \left( {1\over \ell_\Lambda^2} - {\sigma_+ \over 3} \right)}\,, \\
 \label{branches-}
 1+2r_c^{(-)}\mathcal{H}_- = \pm \sqrt{1-4r_c^{(-)\,2} 
 \left( {1\over \ell_\Lambda^2} - {\sigma_- \over 3} \right)}\,,
\end{eqnarray} 
when 
$(r_c^{(\pm)})^{-2}-4\left(\ell_\Lambda^{-2} - {\sigma_\pm
/3}\right)>0$. 
As is discussed above, 
the solution that takes the positive sign 
on both branes corresponds to the normal branch. 
If either of the above solutions takes negative sign, 
the whole setup is in the self-accelerating branch. 
If we have chosen that ${\cal H}$ 
is always positive, the sign in 
Eq.~\eqref{branches-} is guaranteed to be positive by assumption. 
The above two solutions \eqref{branches+} degenerate to one double root when 
$(r_c^{(\pm)})^{-2}-4\left(\ell_\Lambda^{-2} - {\sigma_\pm /3}\right)=0$, 
and consequently we have $1\mp 2r_c^{(\pm)}\mathcal{H}_{\pm} = 0$ in this case. 
For $1- 2r_c^{(+)}\mathcal{H}_{+} = 0$ $ 
(1+ 2r_c^{(-)}\mathcal{H}_- = 0 )$, 
the radion strongly couples to the source 
on the (+)-brane ($(-)$-brane) as 
is seen in Eq.~\eqref{beta+} (Eq.~\eqref{beta-}). 
This can be understood as a consequence of the 
perplexity about the branch choice by the system. 

\section{Nonlinear generalization}
We extend the above method, 
with which we succeeded in reproducing a healthy action 
for bigravity with a scalar corresponding to the radion  
as a low-energy effective theory concerning the perturbation around a de Sitter spacetime, 
to the nonlinear perturbation, although the gradient expansion is not extended to the higher order.
Here we assume that 
\begin{eqnarray}
 \epsilon:=|K\tilde{y}^+|\,,
\end{eqnarray} 
is small as before with 
$g^{(\pm)}_{\mu\nu} \sim \mathcal{O}(1)$, 
$\Delta g_{\mu\nu}:=g^{(+)}_{\mu\nu}-g^{(-)}_{\mu\nu} \sim
\mathcal{O}\left( \epsilon \right)$, 
and $\nabla^2 \sim  \ell_\Lambda^{-2} \sim \mathcal{O}\left( |K|^2/\epsilon\right)$.
Furthermore, the Hamiltonian constraint \eqref{HconstEq} 
and the junction conditions \eqref{junction} 
imply Eq.~\eqref{Rmunu}, as is discussed in Sec.~3. 
As in Sec.~3, we obtain 
$\bar{K}_{\mu\nu}$ and $\bar{g}_{\mu\nu}$ from Eq.~\eqref{expand} as
\begin{eqnarray}
 \bar{K}_{\mu\nu} &=& {\Delta g_{\mu\nu} \over 4\Phi}   \,, \\
 \label{nlbarg}
 \bar{g}_{\mu\nu} &=&  \tilde{g}_{\mu\nu}+\Phi \nabla_\mu \nabla_\nu \Phi 
 + {\Phi^2 \over \ell_\Lambda^2}g_{\mu\nu} 
 + \mathcal{O}\left( \epsilon^2 \right)\,,
\end{eqnarray}
where 
\begin{eqnarray}
\Phi &:=&Ny_0\,, \\
\tilde{g}_{\mu\nu}&:=&{g^{(+)}_{\mu\nu}+g^{(-)}_{\mu\nu} \over 2}\,.
\end{eqnarray}
The Hamiltonian constraint \eqref{HconstEq} becomes
\begin{eqnarray}\label{Ham}
\bar{R}-{12\over \ell_\Lambda^2} = { \Delta g^2 - \Delta g_{\mu\nu}\Delta g^{\mu\nu} \over 16\Phi^2}\,,
\end{eqnarray}
where the tensor indices are raised by $\tilde{g}_{\mu\nu}$.
The bulk action is given as 
\begin{eqnarray}
 S_b &=&{M_{pl}^2 \over2} {1\over2r_c^{(+)}} \oint \sqrt{-g}d^4x N dy 
 \left( {}^5\!R - {12\over \ell_\Lambda^2} \right) \nonumber \\
 &=& {M_{pl}^2 \over2} {1\over r_c^{(+)}} \int d^4x N \int^{y_0}_{-y_0} dy
 \left[ \sqrt{-\bar{g}}\left( \bar{R}-{12\over \ell_\Lambda^2} \right) 
 +  y\overline{{\delta \left( \sqrt{-g}\left( R-{12\over \ell_\Lambda^2} \right) \right) \over \delta g_{\mu\nu}}} \overline{ \partial_y g_{\mu\nu}} \right. \nonumber \\
 &&\left. \qquad\qquad\qquad\qquad\qquad\qquad\qquad\qquad\qquad
 +{y^2\over2} \overline{\partial_y \left( {\delta \left( \sqrt{-g}\left( R-{12\over \ell_\Lambda^2} \right) \right) \over \delta g_{\mu\nu}} \partial_y g_{\mu\nu} \right) } +\cdots \right] \nonumber \\
 &=& {M_{pl}^2 \over2} {1\over r_c^{(+)}} \int d^4x N 
 \left[2y_0 \sqrt{-\bar{g}}\left( \bar{R}-{12\over \ell_\Lambda^2} \right) 
  \right. \nonumber \\
  &&\left. \qquad\qquad\qquad\qquad\quad
  -{2y_0^3\over3}  \overline{{\delta \left( \sqrt{-g}\left( R-{12\over \ell_\Lambda^2} \right) \right) \over \delta g_{\mu\nu}}}
 N^2\left( {1\over \Phi}\tilde{\nabla}_\mu \tilde{\nabla}_\nu \Phi 
 + {1\over \ell_\Lambda^2}\tilde{g}_{\mu\nu} \right)
 \right. \nonumber \\
  &&\left. \qquad\qquad\qquad\qquad\qquad
 + \mathcal{O}\left( |K| \epsilon^2\right) \right] \,, 
\end{eqnarray}
where we use the Hamiltonian constraint \eqref{HconstEq} in the second equality. 
In the third equality, the linear term in $y$ in the integral is integrated to be zero, 
the term in which differentiation with respect to $y$ operates on 
${{\delta \left( \sqrt{-g}\left( R-{12 \ell_\Lambda^{-2}} \right)
\right)/ \delta g_{\mu\nu}}}$ 
becomes higher order of the gradient expansion, 
and we use Eq.~\eqref{KdotEq} and 
$\tilde{R}_{\mu\nu}={3 \ell_\Lambda^{-2}}\tilde{g}_{\mu\nu}
+\mathcal{O}\left(|K|^2\right)$. 
The operator 
$\overline{{\delta \left( \sqrt{-g}\left( R-{12\ell_\Lambda^{-2}} \right) \right) /\delta g_{\mu\nu}}}$ is obtained as 
\begin{eqnarray}
 \int d^4x \overline{{\delta \left( \sqrt{-g}\left( R-{12\over \ell_\Lambda^2} \right) \right) \over \delta g_{\mu\nu}}} h_{\mu\nu}
 =  \int d^4x \left( \sqrt{-\tilde{g}}\tilde{\mathcal{L}}^{\mu\nu}h_{\mu\nu} + \mathcal{O}\left( |K|^2 \right)\right)\,, 
\end{eqnarray}
for an arbitrary tensor $h_{\mu\nu}$. 
Here we defined $\tilde{\mathcal{L}}^{\mu\nu}$ as
\begin{eqnarray}
 \tilde{\mathcal{L}}^{\mu\nu}:= \tilde{\nabla}^\mu \tilde{\nabla}^\nu 
 - \left(\tilde{\Box} + {3\over \ell_\Lambda^2} \right)\tilde{g}^{\mu\nu}\,, 
\end{eqnarray}
where $\tilde{\nabla}$ is the covariant differentiation associated with $\tilde{g}_{\mu\nu}$. 
Using the Hamiltonian constraint Eq.~\eqref{Ham}, the bulk action becomes
\begin{eqnarray}
 S_b &=& {M_{pl}^2 \over2} {2\over r_c^{(+)}} \int d^4x \sqrt{-\tilde{g}}
 \left[ {\Delta g^2 - \Delta g_{\mu\nu}\Delta g^{\mu\nu} \over 16\Phi} 
 +{\Phi \over3} \tilde{\mathcal{L}}^{\mu\nu}
 \left(  \left(\tilde{\nabla}_\mu \Phi \right)\left( \tilde{\nabla}_\nu \Phi \right)
 - {\Phi^2 \over \ell_\Lambda^2}\tilde{g}_{\mu\nu} \right) \right. \nonumber \\
 && \left. \qquad\qquad\qquad\qquad\qquad\quad
  + \mathcal{O}\left( |K|\epsilon^2 \right) \biggr] \right. \,,
\end{eqnarray}
where we use 
$\tilde{\mathcal{L}}^{\mu\nu} \tilde{\nabla}_\mu A_\nu = 0$ 
for an arbitrary vector $A_\nu$ at the leading order of the expansion in $\epsilon$.
The analysis in the linear regime suggests that 
in order to examine the structure of the interaction between $g^{(+)}_{\mu\nu}$ and $g^{(-)}_{\mu\nu}$, 
it is convenient to treat the radion $\Phi$ as an independent variable.
Therefore we consider 
\begin{eqnarray}\label{actionwtL}
 S_b &=& {M_{pl}^2 \over2} {2\over r_c^{(+)}} \int d^4x \sqrt{-\tilde{g}}
 \left[ {\Delta g^2 - \Delta g_{\mu\nu}\Delta g^{\mu\nu} \over 16\Phi} 
 +{\Phi \over3} \tilde{\mathcal{L}}^{\mu\nu}
 \left( \left(\tilde{\nabla}_\mu \Phi \right)\left( \tilde{\nabla}_\nu \Phi \right)
 - {\Phi^2 \over \ell_\Lambda^2}\tilde{g}_{\mu\nu} \right)  \right. \nonumber \\
 &&\left. \qquad\qquad\qquad\qquad\qquad\quad
 + \lambda \left( \bar{R}-{12\over \ell_\Lambda^2} - { \Delta g^2 - \Delta g_{\mu\nu}\Delta g^{\mu\nu} \over 16\Phi^2} \right) \right] \,,
\end{eqnarray}
by imposing the Hamiltonian constraint with a Lagrange multiplier $\lambda$, 
and take the variation with respect to $\Phi$ to eliminate $\lambda$ using the equation of motion for $\Phi$.
$\bar{R}-{12\ell_\Lambda^{-2}}$ is computed as 
\begin{eqnarray}
 \bar{R}-{12\over \ell_\Lambda^2} = \tilde{R}-{12\over \ell_\Lambda^2} +
 \tilde{\mathcal{L}}^{\mu\nu} \left( \Phi \tilde{\nabla}_\mu \tilde{\nabla}_\nu \Phi 
 + {\Phi^2 \over \ell_\Lambda^2}\tilde{g}_{\mu\nu} \right)
 + \mathcal{O}\left( |K| \epsilon \right)\,.
\end{eqnarray}
By taking the variation of Eq.~\eqref{actionwtL} with respect to $\Phi$ and 
using the identity 
$\tilde{\mathcal{L}}_{\mu\nu} \tilde{\nabla}^\mu\tilde{\nabla}^\nu \Psi = 
\tilde{\nabla}^\mu\tilde{\nabla}^\nu \tilde{\mathcal{L}}_{\mu\nu} \Psi =0$ 
for an arbitrary scalar $\Psi$ at the leading order of the expansion in $\epsilon$, 
the equation of motion for $\Phi$ is given in an extremely simple form as
\begin{eqnarray}\label{constraintlambda}
 2 \mathcal{L}_{\lambda} \lambda = \mathcal{L}_{\lambda} \Phi \,,
\end{eqnarray}
where
\begin{eqnarray}
 \mathcal{L}_{\lambda} 
 := \left( \tilde{\nabla}_\mu \tilde{\nabla}_\nu \Phi 
 + {\Phi \over \ell_\Lambda^2} \tilde{g}_{\mu\nu} \right)\tilde{\mathcal{L}}^{\mu\nu} 
 + { \Delta g^2 - \Delta g_{\mu\nu}\Delta g^{\mu\nu} \over 16\Phi^3}\,.
\end{eqnarray}
We ignore the homogeneous solution $\mathcal{L}_\lambda \lambda=0$
by treating $\mathcal{L}_\lambda^{-1}$ as a local operator as in the case of 
$({1-\hat{H}^2\hat{\Box}})^{-1}$ in Sec.~3.
The condition that allows such expansion depends on $\Phi$ and is
difficult to express explicitly. 
Hence, here we just assume that the energy scale is low enough to satisfy this condition. 
Substituting $\lambda={\Phi \over2}$ into Eq.~\eqref{actionwtL}, 
we obtain the 4-dimensional effective action from the bulk action as
\begin{eqnarray}\label{nonlinearaction}
 S_b &=& {M_{pl}^2 \over2} {2\over r_c^{(+)}} \int d^4x \sqrt{-\tilde{g}}
 \left[ {\Delta g^2 - \Delta g_{\mu\nu}\Delta g^{\mu\nu} \over 32\Phi} 
 +{\Phi \over2}\left(  \tilde{R}-{12\over \ell_\Lambda^2} \right) \right. \nonumber \\
 &&\left. \qquad\qquad\qquad\qquad\qquad\quad
 -{\Phi \over6} \tilde{\mathcal{L}}^{\mu\nu}
 \left( \left(\tilde{\nabla}_\mu \Phi \right)\left( \tilde{\nabla}_\nu \Phi \right) 
 - {\Phi^2 \over \ell_\Lambda^2}\tilde{g}_{\mu\nu} \right)  \right]  
+\mathcal{O}(\epsilon^2) 
\nonumber \\
&=& {M_{pl}^2 \over2} {2\over r_c^{(+)}} \int d^4x \sqrt{-g}
 \left[  {\Delta g^2 - \Delta g_{\mu\nu}\Delta g^{\mu\nu} \over 32\Phi} 
 -{1 \over 2\ell_\Lambda^2} \Phi^2 \left( \Box + {4\over \ell_\Lambda^2}\right)\Phi 
 +{\Phi \over2}\left(  \tilde{R}-{12\over \ell_\Lambda^2} \right)
 \right.
 \nonumber \\
&&\left. \qquad\qquad\qquad\qquad\qquad\!
-{1\over6} \left(\nabla_\mu \Phi \right)\left( \nabla_\nu \Phi \right) 
 \left(\nabla^\mu \nabla^\nu - g^{\mu\nu}\Box - R^{\mu\nu} \right)\Phi \right] +\mathcal{O}(\epsilon^2) \,, 
\end{eqnarray}
where $\nabla$ is the covariant differentiation associated with $g_{\mu\nu}$, 
which is indistinguishable from $g_{\mu\nu}^{(\pm)}$ and $\tilde{g}_{\mu\nu}$ 
at the leading order of gradient expansion and 
we use $R_{\mu\nu}={3 \ell_\Lambda^{-2}}g_{\mu\nu}
+\mathcal{O}\left(|K|^2\right) $ in the second equality. 
The non-derivative 
interaction terms between two metrics takes the Fierz-Pauli form 
$\Delta g^2 - \Delta g_{\mu\nu}\Delta g^{\mu\nu} $, which 
is rewritten in terms of dRGT mass term, 
$V_n:=\epsilon_{\mu_1\cdots \mu_n}^{\nu_1\cdots \nu_n} \mathcal{U}^{\mu_1}_{\nu_1}\cdots \mathcal{U}^{\mu_n}_{\nu_n}$ 
for $\mathcal{U}^\mu_\nu:=\sqrt{g_{(+)}^{\mu\rho} g_{(-)\,\rho\nu}}$, as 
\begin{eqnarray}
 \Delta g^2 - \Delta g_{\mu\nu}\Delta g^{\mu\nu}
 =  {4\over J}  \sum_n c_n V_n \,.
\end{eqnarray}
Here we define $J:=-{1\over2}\left( c_1+4c_2+6c_3\right)$  
and the coefficients $c_n$ are constrained by the condition 
that both $g_{\mu\nu}^{(+)}$ and $g_{\mu\nu}^{(-)}$
recover Minkowski metric and $\Delta g_{\mu\nu}\rightarrow 0$ 
in the limit where the energy densities of matter on the branes and the
brane tensions are sent to zero, 
\begin{eqnarray}
 c_4&=&-{1\over24} \left( c_1+6c_2+18c_3 \right) \,, \nonumber \\
 c_0&=&-3\left( c_1+2c_2+2c_3 \right) \,.
\end{eqnarray}
We wrote the interaction between two gravitons in terms of dRGT ones, 
however, we cannot determine the form of mass interactions at higher order of 
$\mathcal{U}^\mu_\nu-\delta^\mu_\nu \sim \mathcal{O}\left( \epsilon \right)$ 
at the order of the present approximation. 
Absorbing $\Phi \tilde{R}$ term into the induced gravity terms 
by conformal transformations for $g_{\mu\nu}^{(\pm)}$, 
as in Sec.~3,  the action for $\Phi$ in the second line 
in Eq.~\eqref{nonlinearaction} is a cubic Galileon~\cite{Nicolis:2008in, Deffayet:2009wt}, 
and hence 
the whole action $S_b+ S_+ + S_-$ describes
a well-known ghost-free system that contains two interacting metrics and one scalar 
which couples only to one of the metrics~\cite{HR1}.
In order to investigate the coupling of radion as a doubly coupled matter, 
we have to step into the higher order of the gradient expansion. 
At the higher order of the gradient expansion, however, higher
derivative terms will appear 
and hence the system may inevitably contain extra degrees of freedom 
corresponding to the other massive KK gravitons. 
Therefore, it may not be allowed to obtain doubly coupled matter model using our method, 
because at the leading order of the gradient expansion the couplings to
two metrics is indistinguishable,  
and 
the radion may couple to more than two tensor modes 
if we expand to higher order. 

\section{Summary}
In this paper, we intended to obtain the ghost-free bigravity action 
with a single scalar field 
from DGP 2-brane model with an unstabilized radion, 
by solving the bulk equations for given boundary metrics $g_{\mu\nu}^{(\pm)}$ 
at $y=\mp y_0$ and integrating out the bulk degrees of freedom 
under the approximation that $g_{\mu\nu}^{(+)}-g_{\mu\nu}^{(-)}\sim 
|KNy_0| \ll1$.
As a result, we obtained an 
action written in terms of $g_{\mu\nu}^{(\pm)}$ 
as a 4-dimensional effective theory of DGP 2-brane model, 
which is reduced to a healthy bigravity model with a scalar field, 
as expected. 
Truncating the result at the leading order of the gradient expansion, 
we obtained the Fierz-Pauli quadratic mass term as the interaction between two gravitons, 
though we worked on non-linear perturbation. 
The scalar field couples to only one of the metrics $g_{\mu\nu}^{(\pm)}$ and 
its equation of motion does not contain higher-order time derivative. 
To realize such a setup of the model, 
we need to tune the brane tension, so that $|K|$ is 
sufficiently small. 
Since this tuning is easily broken by additional matter fields, 
the energy density of the matter fields on the branes must be small
enough. 

We succeeded in obtaining the Fierz-Pauli mass interaction naturally 
from DGP 2-brane model at the lowest order of the gradient expansion.
However, it is difficult to extend this method to the higher order of the gradient expansion
and to obtain the nonlinear dRGT mass terms, 
because it will produce complex and higher-derivative interaction terms.
Such interaction terms seem to lead extra degrees of freedom in addition to 
two gravitons and one scalar radion, 
which will correspond to the appearance of other bulk degrees of freedom. 
It will be possible to investigate only the higher-order mass interactions between two metrics 
by taking the limit $\alpha m_*^2 \rightarrow 0$,
although in this case the mass interaction obtained from the bulk action 
should be different from the dRGT one.
This is because the self-accelerating branch is chosen and radion ghost or Higuchi ghost appears. 
If ghost appears, the interaction between two metrics $g_{\mu\nu}^{(\pm)}$ 
will not take the dRGT form, as is shown in Sec.~3.
However, it might be suggestive that the Fierz-Pauli mass term 
was recovered by fixing the radion by hand in Sec.~3. 
The investigation of the higher-order mass interactions
by considering more extended models is left for future work.

\acknowledgments
We would like to thank Claudia de Rham for useful discussions 
during the Nordita program "Extended Theories of Gravity". 
YY is supported by the Grant-in-Aid for Japan
Society for the Promotion of Science(JSPS) Fellows No. 15J02795
and TT is supported by the Grant-in-Aid for Scientific
Research (Nos. 24103006, 24103001, 26287044 and 15H02087).

\end{document}